\documentclass{PoS}
\usepackage{rotating}

\title{Flows and Shocks: Some Recent Developments in Symbiotic Star and Nova
Research}

\ShortTitle{Flows and Shocks in Symbiotic Stars and Novae}

\author{\speaker{J. L. Sokoloski}\\
        Columbia Astrophysics Laboratory\\
        E-mail: \email{jeno@astro.columbia.edu}}

\author{Stephen Lawrence\\
        Hofstra University\\
        E-mail: \email{Stephen.S.Lawrence@hofstra.edu}}

\author{Arlin P. S. Crotts\\
        Columbia Department of Astronomy\\
}

\author{Koji Mukai\\
        Goddard Space Flight Center and University of Maryland Baltimore County\\
        E-mail: \email{Koji.Mukai@nasa.gov}}

\abstract{There have been several surprising developments in our
  understanding of symbiotic binary stars and nova eruptions over the
  last decade or so based on multiwavelength data. For example,
  symbiotic stars without shell burning have been revealed through
  their X-ray emission, UV excess, and UV variability. These purely
  accretion powered symbiotic stars have much weaker optical emission
  lines and radio emission than those with shell burning, and therefor
  harder to discover, yet may be as numerous as the burning symbiotic
  stars. Interestingly, both types of symbiotic stars are capable of
  driving strong outflows, leading to colliding wind X-ray emission
  and spatially resolved X-ray jets. For nova eruptions, the most
  surprising discovery has been that they are capable of particle
  acceleration as evidenced by Fermi detection of novae as transient
  GeV gamma-ray sources. For nova eruptions in cataclysmic variables,
  this implicates internal shocks, between a slow, dense outflow and a
  fast outflow or wind. Other signatures of shocks include thermal
  X-rays and non-thermal radio emissions, and a substantial fraction
  of optical emission may be shock-powered in the early phase of
  novae. Radio (V959 Mon) and HST (V959 Mon and T Pyx) images of nova
  shells within a few years of their respective eruptions suggest that
  nova ejecta may commonly consist of an equatorial ring and a bipolar
  outflow.}

\FullConference{Accretion Processes in Cosmic Sources --- APCS2016 ---\\
		5-10 September 2016,\\
		Saint Petersburg, Russia}

\newcommand{\msun}{{M}_{\odot}}
\newcommand{\lsun}{{L}_{\odot}}
\newcommand{\lwd}{{L}_{\rm WD}}

\newcommand{\msyr}{\msun\; {\rm yr}^{-1}}
\newcommand{\mdotwd}{\dot{M}_{\rm WD}}

\begin{document}

\section{Introduction: shocks point the way in symbiotic stars and novae}

During the past 5 to 10 years, multiwavelength observations and
theoretical studies of symbiotic binary stars and nova eruptions have
led to some surprising developments in these fields.  In both
symbiotic binaries and novae, the key findings are related, at least
in part, to flows, shocks, and the high-energy emission that results.

Symbiotic stars are wide, interacting binaries in which a compact
object accretes from a red-giant (RG) companion.  Orbital periods
range from years to decades, and binary separations range from AU to
tens or even hundreds of AU.  Here, we restrict our discussion to
symbiotics that contain white dwarf (WD) accretors.  In some such
symbiotic stars, the interaction is powered solely by the release of
gravitational potential energy as matter falls onto the WDs.  In
others, the temperature and luminosity of the WD indicate that the
interaction is powered primarily by the quasi-steady burning of
hydrogen-rich fuel in a shell on the WD surface (see \cite{murset1991}
for a compilation of WD temperatures and luminosities).  The degree to
which this shell burning is: 1) residual burning from past novae
(which perhaps lasts longer in wide binaries than in cataclysmic
variables; CVs); 2) the result of accretion at a high rate; or 3) a
combination of both is not yet evident.  But recent observations at
radio through X-ray wavelengths reveal that the presence or absence of
shell burning on the WD determines the characteristics of a symbiotic
at every waveband.  X-rays are particularly valuable for diagnosing
the inner accretion flow.  Lastly, symbiotics without shell burning, which
have been difficult to identify in the optical, are worth searching
for in the X-rays and the ultraviolet (UV) time domain.  This
heretofore hidden population offers the opportunity to investigate
wind-fed accretion disks with radii of $10^{12}$ to $10^{13}$~cm,
their jets, and the evolution and statistics of wide, interacting
binaries --- including the production of type Ia supernovae.

Nova eruptions occur on accreting WDs in wide, symbiotic binaries and
in the much tighter cataclysmic variables (CVs).  Between 2010 and
2014, the $Fermi$ Gamma-Ray Space Telescope astonished the
astronomical community with its finding that novae constitute a new
class of GeV $\gamma$-ray sources (e.g.,
\cite{abdo2010,ackermann2014}).  It turns out that nova events
generate $\gamma$-rays both when the erupting WD is embedded within
the wind of a red-giant companion, and when it is not.  As we will
discuss below, the hunt for the origin of these $\gamma$-rays has
exposed the degree to which the outflows from novae consist of
multiple, distinct flows, and the crucial role of powerful shocks in
almost all aspects of novae.

\section{Symbiotic stars with and without shell burning}

{\bf Energetics.}  The luminosity of an accreting WD with a red-giant
companion is set chiefly by whether or not accreted material burns
quasi-steadily in a shell on the surface of the WD.  If material in
the WD envelope is burning at approximately the rate given by the core
mass-luminosity relation \cite{paczynski1978}, then the
luminosity of the WD is given by $L_{\rm WD}({\rm with\, burning}) 
\sim 10^3\, \lsun$.  If, on the other hand, shell
  burning is absent, the luminosity of the WD is roughly that
  generated by accretion,
\begin{eqnarray}
L_{\rm WD}({\rm without\; burning}) & \approx & G M_{\rm WD} \mdotwd /
      R_{\rm WD} \\
 & \approx & 10\, \lsun \left(\frac{M_{\rm WD}}{0.6\, \msun} \right)
      \left( \frac{\mdotwd}{10^{-8}\, \msyr} \right) \left(
      \frac{R_{\rm WD}}{10^9\, {\rm cm}} \right)^{-1},
\end{eqnarray}
\noindent
where $M_{\rm WD}$ is the mass of the WD, $\mdotwd$ is the rate of
accretion onto the WD, and $R_{\rm WD}$ is the radius of the WD.
As we will see below, it is not just the WD luminosities that distinguish
WD-plus-RG interacting binaries with and without shell
burning; the two different sources of power for the WD create
observational differences across the electromagnetic
spectrum.

\begin{figure}
\begin{center}
\includegraphics[width=0.8\textwidth]{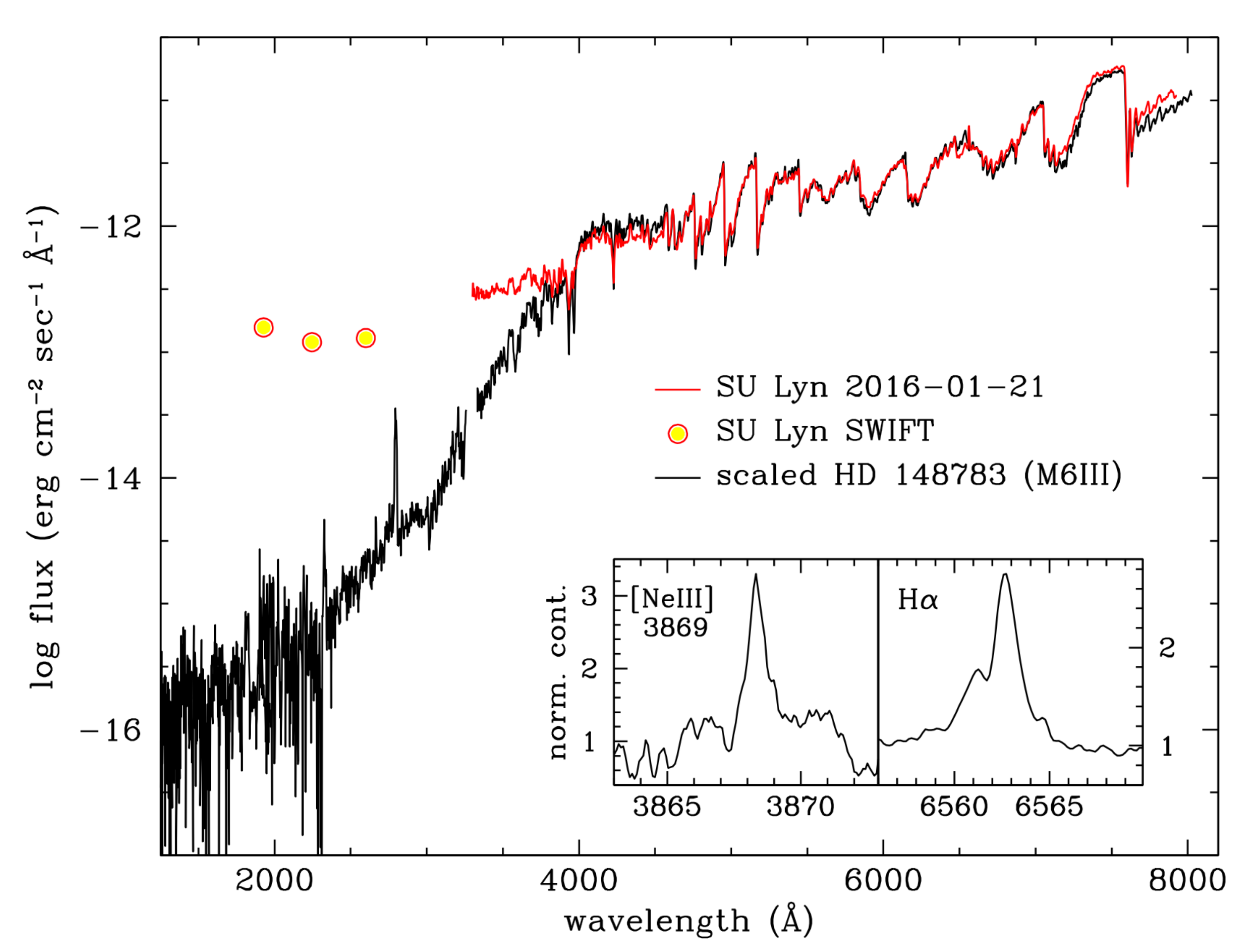}
\end{center}
\caption{UV through optical spectrum of non-burning symbiotic star
  SU~Lyn, reproduced from \cite{mukai2016}.  The yellow UV points for
  SU~Lyn are 
  from $Swift$/UVOT.  The black curve shows $IUE$ and
  optical spectra of a standard MIII star, for comparison.  Longward of about
  4000$\AA$, the low-resolution optical spectrum of SU~Lyn is
  difficult to distinguish from that of a normal red giant.
  The insets show faint emission lines (of H$\alpha$ and
  [Ne~III]~3869$\AA$) detected with a high-resolution 
  optical spectrum. SU~Lyn shows how difficult it would be to detect
  such systems with low-resolution optical spectroscopy alone.}
\label{fig:sulyn}
\end{figure}

{\bf Optical spectra.}
Because symbiotic stars have traditionally been defined and identified
by their optical spectra, it is worth starting our discussion with the
impact of shell burning (or the lack thereof) on the optical emission.
In a CV, the accretion disk usually dominates
the optical emission whether or not shell burning is present, because
most light from the WD photosphere is radiated shortward of the
optical.
In an WD-plus-RG
interacting binary, on the other hand, the donor is quite bright, and
the nebula reprocesses some of 
the UV to soft X-ray emission from the burning WD into the optical.  Thus,
if the WD is hot and luminous due to shell burning, the optical emission is
frequently dominated by light from the red giant donor and/or
the ionized nebula \cite{skopal2005}.  In particular, the
strength of the optical line and recombination-continuum emission from
the nebula depends on the luminosity (and to some extent
temperature)
of the accreting WD, which in turn depends mainly upon whether
hydrogen-rich fuel is burning quasi-steadily on its surface.  With
ongoing burning, the temperature of the WD sits above $\sim10^5$~K,
and the luminous WD ionizes a large portion of the nebula.  The nebula
then generates the strong, high-ionization state optical emission
lines that are the hallmark of classical symbiotic stars (e.g.,
\cite{kenyon1986,munari2002,mska2003} and references therein).  Without shell
burning on the WD, a predominantly neutral --- and therefore optically
faint --- nebula allows photospheric emission from the red giant, and
in some cases the accretion disk to dominate the optical spectrum,
with only weak, if any, emission lines. Extreme examples of such
systems include the symbiotic recurrent novae (such as T~CrB in its
normal, quiet state); the
well-known jet-producing binaries CH~Cyg, R~Aqr, and MWC~560; and
systems with very hard X-ray emission such as RT~Cru, V648~Car, and
SU~Lyn (Fig.~\ref{fig:sulyn} shows the UV to optical spectrum of
SU~Lyn, from \cite{mukai2016}).  WD-plus-RG interacting binaries in
this state have sometimes been referred to as {\em weakly symbiotic}
or {\em symbiotic-like} binaries.  WD-plus-RG binaries without shell
burning, however, are not necessarily weakly interacting.

{\bf What's in a name?}  Although WD-plus-RG interacting binaries
without quasi-steady shell burning on the WD often
lack the strong optical
emission lines that motivated the original definition of symbiotic
stars,
we argue that it is nevertheless appropriate
and useful to include such systems among the so-called symbiotic
stars\footnote{Though a bit counter-intuitive to extend the definition
of symbiotic stars to objects that do not show much of the optically
defined ``symbiotic phenomenon" \cite{mska1988}, we
hope readers will agree that it is more
natural than introducing a new name for a what is almost certainly a
transient state.  Additionally, 
the name {\em WD-plus-RG interacting binaries} is cumbersome and obscures the
connection to past research on symbiotic stars.}.  As in past work
(e.g., \cite{kennea2009,luna2013,mukai2016}), we therefore
define a WD symbiotic system as {\em a binary in
which a red giant transfers enough material to a WD for the
interaction to produce an
observable signal at some 
waveband}.  
Considering WD-plus-RG binaries with and without shell burning
together is meaningful in part because the rates of accretion onto the
WDs in systems with and without shell burning can be comparable ($\sim
10^{-10}$ -- $10^{-7} \msyr$ (e.g.,
\cite{hachisu2001,sokoloski2010,eze2010, luna2013}.
For WD-plus-RG
binaries without shell burning, however, the interaction is normally
more apparent in X-ray spectra and optical-UV fast photometry than in
optical spectra (see below).  In fact, at X-ray
wavelengths and in optical-UV fast photometry, signatures of
binary interaction can be {\em stronger} in non-burning than in
burning symbiotics.
Finally, because shell
burning on the WD is almost certainly a transient phenomenon, each
WD-plus-RG binary is likely to spend some time with and some time
without shell burning.  Shell burning may even turn on and off more
than once.  We thus
refer to WD-plus-RG binaries with and without shell burning on the surface
of the WD as {\em burning} and {\em non-burning} symbiotics,
respectively.

{\bf X-rays: revealing the inner accretion flow.}  Looking beyond the
optical spectrum, the nature of X-rays from a symbiotic star is also
dictated in large part by whether or not quasi-steady shell burning is
present on the surface of the WD.  For example, some non-burning
symbiotics emit highly absorbed X-rays with energies greater than
several keV and as high as tens of keV (which we refer to as {\em
  hard} X-rays).  When such binaries are nearby, their hard X-ray
emission can be bright enough to reveal flickering-type variations on
time scales of minutes to hours, and spectra that are well modeled as
isobaric cooling flows (as in \cite{mukai2003}).  This X-ray spectral
component was referred to by \cite{luna2013} as $\delta$-type X-ray
emission, and it most likely emanates from an accretion-disk boundary
layer (e.g., \cite{luna2007,kennea2009, eze2010, luna2013, mukai2016,
  ilkiewicz2016}).
Evidence for a
boundary-layer origin includes: 1) rapid, stochastic brightness
variations like those for the boundary layers of CVs;
2)
cooling flow spectra, like those expected from flows onto the surface
of a WD (and observed in CVs (e.g., \cite{pandel2005}); 3) high intrinsic,
partial-covering absorption, indicating that the source of hard X-rays
is small and located behind the nebular and/or disk wind; and 4) a
lack of detectable modulation of the X-ray brightness at the WD spin
period,
indicating that the X-rays are unlikely to emanate from magnetic
accretion columns(see \cite{mukai2017} for a review of X-ray properties of
accreting WDs).
Although \cite{ducci2016} have suggested magnetic accretion as the
origin of hard X-rays in at least one non-burning symbiotic (RT~Cru),
we 
contend that
the lack of spin modulations in the hard X-ray emission
from all non-burning symbiotics supports the boundary-layer
interpretation for the class as a whole.
In burning symbiotics, 
on the other hand, 
hard X-ray emission, with energy greater than a few keV,
seems to be absent, perhaps
because the high flux of FUV photons from the luminous, hot WDs
Compton-cools the boundary layers out of the X-ray regime (as
described in general terms in \cite{frank2002}).
So, whereas searches for red giants with strong,
high-ionization state optical emission lines
      preferentially uncover burning symbiotics, searches for red giants
with hard X-ray emission would preferentially (or exclusively)
uncover symbiotics without shell burning.

\begin{figure}
\begin{center}
\includegraphics[width=0.45\textwidth]{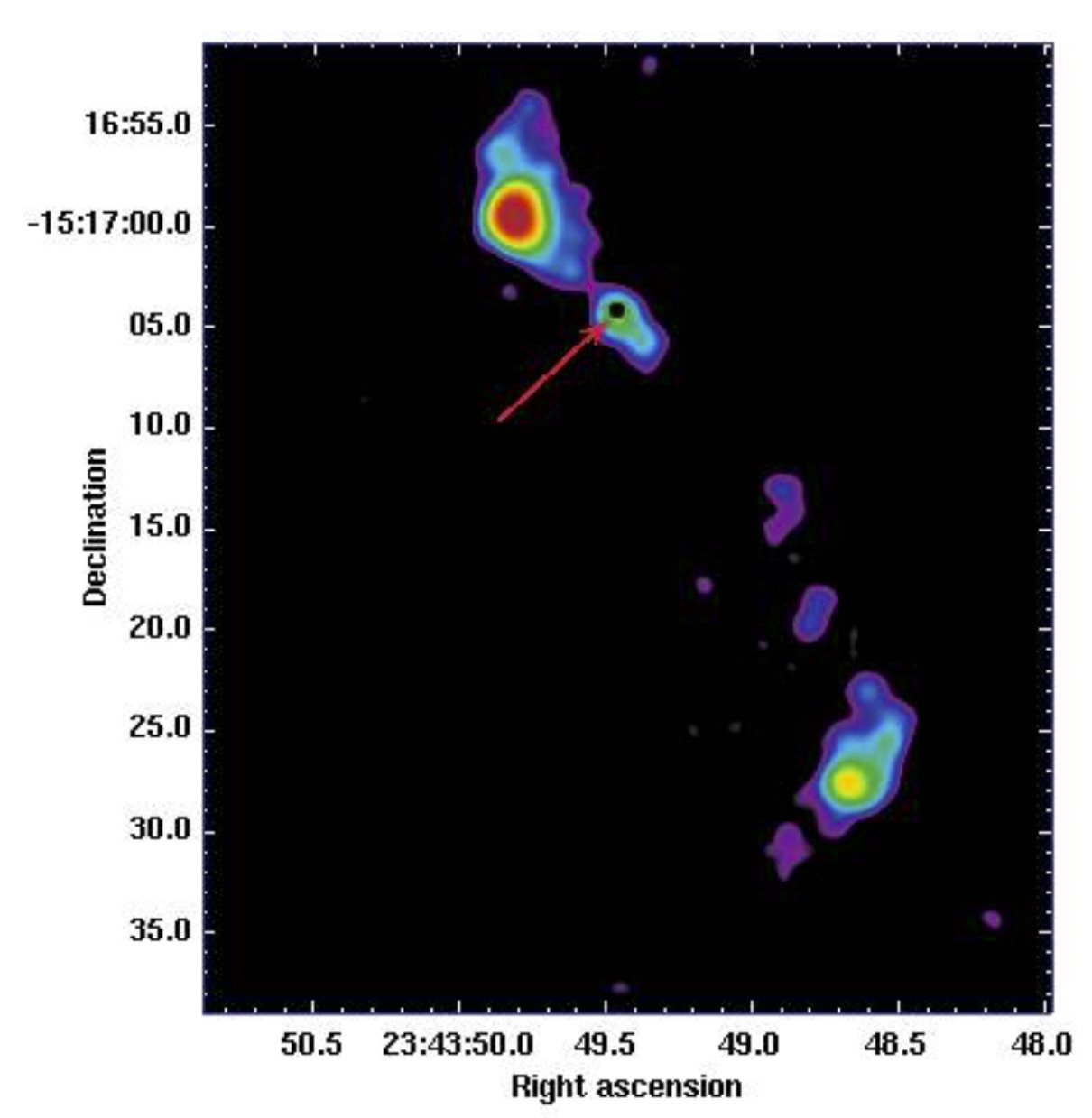}
\includegraphics[width=0.44\textwidth]{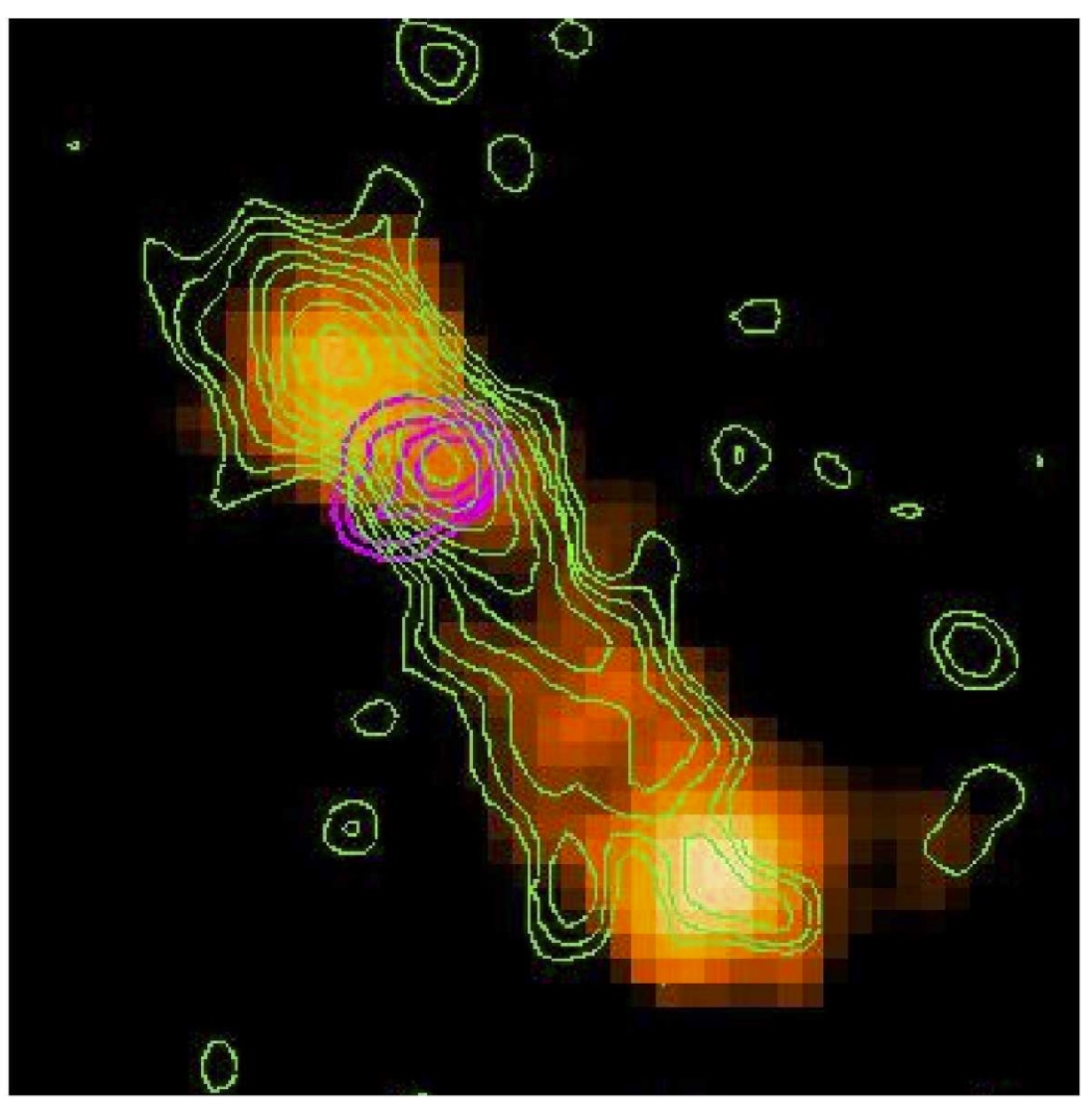}
\end{center}
\caption{$Chandra$ images of the X-ray jets from non-burning
  symbiotics R~Aqr (left) and CH~Cyg (right), reproduced from
  \cite{kellogg2007} and \cite{karovska2010}. In the R~Aqr image, the
  0.2 to 3.5~keV X-rays were smoothed, and the red arrow indicates the
  location of the central binary.  In the CH~Cyg image, green contours
show the radio flux density at 5~GHz, magenta contours show the hard,
6-7~keV X-rays, and the yellow and orange image shows the location and
strength of soft X-rays.}
\label{fig:xjets}
\end{figure}

{\bf Softer X-rays from shell burning, jets, and colliding
winds.} 
If a WD with shell burning is massive enough for photospheric
temperatures to reach at least several hundred thousand degrees,
burning
symbiotics emit supersoft X-ray emission, which \cite{murset1997} dubbed
$\alpha$-type X-ray emission.  Examples of symbiotics with such
emission include AG~Dra, RR~Tel, and StH$\alpha$~32
\cite{murset1997,orio2007}. 
Such
supersoft X-rays are only  detectable if the column of absorbing
material is fairly low.
Both burning and non-burning symbiotics are able to
produce so-called $\beta$-type thermal X-ray emission, 
in which most photons have energies less than approximately 2 or 3~keV.
The lower column of absorbing material compared to $\delta$-type
X-ray emission, the lack of minute-to-hour time scale
variability, and the lower plasma temperatures than typically found
in boundary layers all
suggest that it arises from colliding winds
\cite{murset1997,luna2013,nunez2016} or within collimated jets (e.g.,
\cite{kellogg2001,galloway2004,kellogg2007,nichols2007,karovska2010,stute2011}.
R~Aqr and CH~Cyg provide that most dramatic examples of X-ray emission
from shock-heated plasma within WD jets (Fig.~\ref{fig:xjets}).  For
sources that are faint in the X-rays, it is sometimes difficult to
distinguish between $\beta$-type and $\delta$-type X-ray emission
\cite{luna2013}. But even if it is not always clear whether moderately
hard 
X-rays come from colliding winds, a jet, an accretion-disk boundary
layer, or magnetic accretion columns, observations with $Swift$,
INTEGRAL, $Chandra$, and most recently, $NuSTAR$ have certainly made
it clear that the X-ray emission from symbiotic stars is complex,
variable, both soft and hard, and dependent upon the presence or
absence of shell burning.

{\bf Disk flickering in the UV.} 
Accretion in CVs, X-ray binaries, and active galactic nuclei leads to
stochastic brightness variations on time scales from roughly the
dynamical time near the inner edge of the disk to viscous time scales
throughout the disk.
UV light curves from
$Swift$/UVOT have now shown that --- as long as no shell burning is
present to hide the disk --- symbiotic stars fit this same
pattern.  The classic accretion signature of large-amplitude
flickering
(fractional rms amplitude greater than about 10\%) 
was not initially found to be
pervasive in symbiotic stars because early searches were performed in the
optical (e.g., \cite{sokoloski2001,gromadzki2006}).  Although many
symbiotics probably 
contain accretion disks whose inner regions are similar to disks in
CVs \cite{livio1984,mohamed2007,devalborro2009},
the disk is 
not usually a major contributor to the optical light.  Moving to the
UV, where the red giant makes a much smaller contribution, enabled
\cite{luna2013} to detect disk flickering from a larger fraction of
symbiotics than \cite{sokoloski2001} did in the optical.  Moreover,
based on a comparison between UV variability and X-ray spectral
properties, they concluded that symbiotics
without shell 
burning had a much greater  propensity to generate large-amplitude UV
flickering. 
In symbiotics with shell burning, UV disk light is outshined by
emission the ionized nebula -- which does not generally vary on time
scales of minutes to hours.  Supporting the idea that flickering from
the accretion flow is easier to detect if the WD does not support
shell burning, all of the symbiotics known to have strong optical
flickering are non-burning symbiotics (e.g., CH~Cyg, V407~Cyg, RT~Cru,
MWC~560, and symbiotic recurrent novae such as RS Oph and T CrB).

{\bf Finding the accretion structures in symbiotic stars.}  Therefore,
by moving beyond the optical band, \cite{luna2013} (and
others)
identified robust techniques for detecting
the accretion disks in at least some symbiotics.  One can probe the
accretion flow by observing in the hard X-rays and/or in the UV time
domain.  Additionally, Fig.~\ref{fig:sulyn} and work by
\cite{sahai2015} show that even without variability information, one
can identify red giants with UV excess as candidate non-burning symbiotics.
 Conversely, these same authors found that seeing a classic
signatures of accretion --- in this case UV or optical disk
flickering, or hard X-ray emission --- furnishes strong evidence that
a given symbiotic is powered by accretion alone rather than shell
burning.  Another consequence of these findings is that there is no
evidence that the inner accretion flows in symbiotics and CVs are
wildly different; accretion is just usually hidden in symbiotics with
shell burning.

{\bf Radio emission from burning vs non-burning symbiotics.}
Completing our examination of the observational signatures of burning
and non-burning symbiotics across the electromagnetic spectrum, clear
disparities exist between the radio brightnesses of the two types of
symbiotics.  In a seminal series of papers,
\cite{seaquist1984, seaquist1990, seaquist1993} found that most of
  the symbiotic stars they 
  detected with the Very Large Array (VLA) 
had radio flux densities on the order of mJy, consistent with
free-free emission from the ionized wind of the red giant.  Because
their target lists consisted primarily of objects with strong optical
emission lines, the majority of the sources they detected were likely
burning symbiotics.  When \cite{westonphd} specifically targeted 11
non-burning symbiotics for radio observations, on the other hand, she
found that about half of these objects had radio flux densities of
approximately $10\,\mu$Jy or less (T~CrB, ER~Del, CD~-27~8661, TX~CVn,
MWC~560, CD~-28~3719, and BD~-21~3873).  The others on her target list
(NQ~Gem, UV~Aur, ZZ~CMi, and Wray~15-1470) had flux densities of a few
tenths of a mJy (but with no indication that these sources are
particularly distant).  The finding that non-burning symbiotics tend
to have very faint radio emission is consistent with the optical
emission lines from these sources being weak, and the ionized regions of the
red-giant winds being small compared to those in burning symbiotics.
In terms of diagnosing the source of the WD's power, a luminosity
of $\lwd \sim 10^3\,\lsun$ or greater is a compelling sign that shell
burning is the source of power; it can, however, be challenging to
measure $lwd$ (especially if UV observations are not available).
That non-burning symbiotics generally have much weaker radio emission
than burning symbiotics furnishes a more easily attainable
observational diagnostic of the burning status.

Fortuitously, the low quiescent-state radio
luminosity of non-burning symbiotics might also allow radio
emission to be used as an effective probe of transient, bipolar
outflows.  For instance, \cite{lucy2017} discovered a major radio
brightening during the 2016 optical high state of MWC~560
\cite{munari2016}. 
 They concluded that the radio brightened as a result of
an increase in 
the power of the well-known (possibly bipolar) outflow in concert 
with the rise in the accretion rate onto the WD.

{\bf Selection bias and the missing population of interacting
  binaries.}  Considering our juxtaposition of burning and  
non-burning symbiotics, it becomes evident that optical spectroscopic
surveys, which excel at finding burning symbiotics, almost certainly
miss many (or even most) non-burning symbiotics.  Recent and
on-going
optical spectroscopic surveys are adding to the
number of symbiotics and candidate symbiotics, both in our galaxy
(e.g., \cite{corradi2008,corradi2010, viironen2009}
and in nearby
galaxies \cite{mska2014}.  But with a bias toward finding burning
symbiotics, these data could tempt us into drawing mistaken
conclusions about, for example, number densities and birth rates of
symbiotic stars, and the distributions of their properties.  For
example, if shell burning is easier to establish and
maintain on low-mass WDs, symbiotics identified by strong lines in
their optical spectra would tend to contain low-mass WDs, or higher
rates of mass transfer.  Indeed, 
compilations of WD mass estimates for optically selected symbiotics do
suggest that this sample on average has low-mass WDs (e.g., \cite{mska2003}).  Conversely, symbiotics with hard X-ray emission are more
likely to contain high-mass WDs (e.g., \cite{luna2007, luna2008,eze2010})
and/or lower 
rates of mass transfer.  In the most extreme case, a reliance on
optical spectroscopy could mean we have
missed almost an entire population of interacting binary stars --- the
non-burning symbiotics.  Work by \cite{morihana2016} suggesting that an
appreciable fraction
of the so-called Galactic ridge X-ray
emission could be due to WDs accreting from red-giant companions; past
detections of X-rays from red giants by \cite{vandenberg2006}); and 
estimates of the space density of non-burning symbiotics by
\cite{mukai2016} all
raise the intriguing possibility that the population of non-burning
symbiotics is significant.  Finally, because nearby
non-burning symbiotic stars power some of the most dramatic WD jets
(e.g., R Aqr, CH Cyg; Fig.~\ref{fig:xjets}),
unveiling the population of
non-burning symbiotics could be particularly exciting for
the study of astrophysical disks and jets.   We are currently laying
the groundwork for moving beyond optical spectroscopy to search for
this missing population.

\section{Novae: colliding flows explain $\gamma$-rays and more}

{\bf Two distinct flows.} Turning to novae, we now direct our
attention from WD inflows to outflows.  At a fundamental level, the
ejecta from 
novae appear to consist of two main components: a slow, dense outflow with a
maximum velocity of less than about $1000$~km~s$^{-1}$ and a
fast outflow or wind with a maximum expansion speed of several thousand
km~s$^{-1}$.  As a nova remnant expands, the slow flow
is observed as a dense core in radio
or optical images (e.g., $HST$ images of V959~Mon and T~Pyx,
below), whereas the fast flow often takes the form of more extended,
bipolar lobes.
In addition to the two novae discussed below, the $\gamma$-ray bright
nova V339~Del affords a nice example of this behavior
\cite{schaeferg2014, gehrz2015}.  Because the rapidly expanding outer
structures tend to fade within a few years,
many of the
optical images of old nova shells (observed more than a decade
after the novae eruption (e.g., \cite{gill1998,gill2000} may record
principally the slow-moving component of the  
  ejecta. 
Supporting this idea,
 these old shells
regularly show expansion speeds (size
    scale divided by time since eruption) of less than
 1000~km~s$^{-1}$.
In terms of optical line profiles, \cite{gehrz2015} 
note that \cite{kolotilov1980,prabhu1987a,prabhu1987b,schaeferg2014}
all conclude that high-velocity wings on emission lines detected
in early spectra of CO novae ``originate in low-density, outlying gas,
and that the lower velocity given by the line core is representative
of the expansion rate of the bulk of the ejecta."  Critically, the
ejecta do not consist of a structure that expands uniformly from
$t_0$ (the time of the thermonuclear runaway; TNR) -- rather, the fast
flow plows through and/or around the slow flow, getting
shaped by it and giving rise to strong shocks.

{\bf Equatorial rings.} Within the slower component of the ejecta, the
density is often enhanced toward the equatorial plane (e.g.,
\cite{paresce1995, chochol1997, eyres2005, sokoloski2013, chomiuk2014}).
This equatorial torus presumably
shapes the faster component of the ejecta into a bipolar morphology.
Evidence for slow, equatorial rings and faster, bipolar
outflows comes from the complex profiles of optical emission
lines (e.g.,
\cite{hutchings1972,solf1983,gill1999,ribeiro2011,shore2013,ribeiro2013}).
Optical and radio imaging 
support the idea that equatorial rings and bipolar
lobes are common (e.g., \cite{gill1998,woudt2009} and the
$HST$ images below).  Minor features such as polar blobs,
polar rings, general clumpiness, and an early `puff' of ejecta are
also often 
observed,
but 
the most energetically
important components of the ejecta appear to be the slow and fast flows,
which tend to produce equatorial rings and bipolar lobes,
respectively.  Furthermore, \cite{chomiuk2014} used radio observations of
the $\gamma$-ray bright nova V959~Mon to infer that collisions
between a dense equatorial torus and a faster flow 
led to shocks that accelerated particles,
explaining the stunning recent discovery that many normal novae
produce GeV $\gamma$-rays \cite{ackermann2014,cheung2016}.

{\bf Generation of $\gamma$-rays.} Moreover, the scenario for $\gamma$-ray production inferred for
V959~Mon by \cite{chomiuk2014} could be broadly applicable.
Relativistic particles that generate $\gamma$-rays from
classical novae (either via inverse-Compton scattering or the
decay of neutral pions) are almost certainly accelerated in shocks.
Although some nova-producing WDs that are embedded in
the wind of a red-giant companion (which we refer to as {\em embedded
novae}) produce $\gamma$-ray emission via external shocks between the
ejecta and the pre-existing circumbinary material
(e.g., in V407~Cyg \cite{abdo2010,nelson2012,orlando2012} and V745~Sco
\cite{cheung2014, orio2015, orlando2017}),
most $\gamma$-ray
bright novae are not embedded in such dense environments
\cite{ackermann2014}.  
Thus, the location and properties of shocks within nova shells are
crucial for understanding how normal (non-embedded) novae generate
$\gamma$-rays.  
With its fortuitously high inclination
\cite{page2013,shore2013b,ribeiro2013}, 
V959~Mon displayed radio synchrotron-producing shocks between a
bipolar flow extending to the east and west, and a less
extended equatorial torus aligned north-south and viewed from the
edge \cite{chomiuk2014}.  But we know that equatorial rings
and shocks are both common in non-embedded novae. As we
discussed above, imaging at various wavelengths and optical
emission-line profiles frequently reveal equatorial
rings.  X-ray emission with energy greater than approximately 1 keV, and radio
synchrotron emission (e.g., \cite{weston2016a,weston2016b}), reveal
shock-heated gas (e.g., \cite{mukai2001}) 
and particles accelerated in shocks, respectively.
Thus, the ingredients that led to gamma-rays in V959~Mon are
common.  Most novae could therefore perhaps generate $\gamma$-rays as in
V959~Mon.

\begin{figure}
\begin{center}
\includegraphics[width=0.95\textwidth]{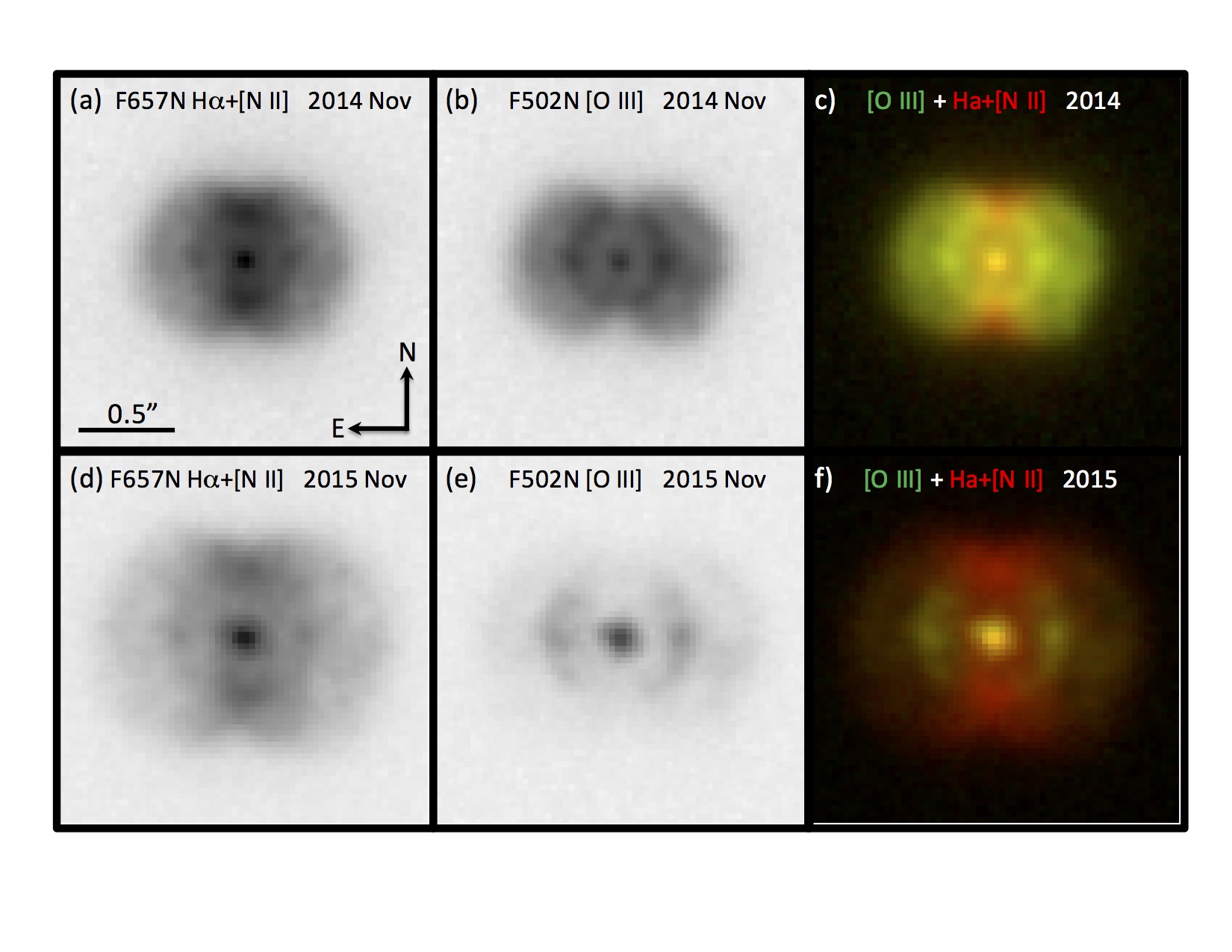}
\end{center}
\caption{$HST$/WFC3 images of V959~Mon, which was
the first
  classical nova to be detected by $Fermi$.  
Images have been drizzled to 0.03$''$ pixel scale. All
   frames are 64 pixels = 1.92$''$ square, N up, with identical logarithmic
   intensity scaling and stretch (pegged to brightest pixel in F657N in
   2014). (a) F657N 2014 November. (b) F502N 2014 November. (c) RGB color
   frame mapping G channel to F502N 2014, R channel to F657N 2014 and B
   channel left blank. (d) F657N 2015 November. (e) F502N 2015 November.
   (f) As with panel (c), but for the 2015 epoch. 
}
\label{fig:v959monhst}
\end{figure}

{\bf $HST$ observations of V959~Mon.} To examine the dominant
structures in the ejecta from V959~Mon more 
clearly, we turn to $HST$ images and imaging spectroscopy.  On 2014
November 21 and then 2015 November 30, HST observed V959~Mon with the
WFC3 camera (and on 19 December 2014 and 21 December 2015 with STIS;
program 13715 [PI: Sokoloski]).  With an initial discovery date for
the nova (in the $\gamma$-rays) of 2012 June 19 (\cite{ackermann2014};
see also \cite{cheung2012}), the $HST$ observations captured the 
state of the remnant approximately 
2.5 and 3.5
years after the
start of the outburst.  Fig.~\ref{fig:v959monhst} shows $HST$/WFC3
images of V959~Mon through the F657N filter (H$\alpha +$[NII]; left
column) and F502N filter ([OIII]; right column).

The [H$\alpha + NII$/F657N] images of V959~Mon are dominated by
H$\alpha$ emission, 
which traces dense, ionized gas.  They show the central
binary (the unresolved central point source), an overall bipolar structure
with a major axis in the east-west direction, and four knots
lying on a circular feature that spans the minor axis (north-to-south)
of the remnant.  The outermost, bipolar shape is roughly consistent
with the morphology that \cite{ribeiro2013} inferred from the profiles of
optical emission lines and that \cite{chomiuk2014} detected in
their VLA images on day 126.  
The STIS spectroscopic image indicates that the eastern lobe is
somewhat blueshifted (tilted slightly toward the observer) and the
western lobe is somewhat redshifted (tilted slightly away from the
observer).  The two strongest knots, along the north-south axis, are
consistent with limb-brightened emission from the edge-on equatorial
torus that dominated the radio images on day 615 \cite{chomiuk2014}.
Additionally, the kinematics of the central circular
structure from our two epochs of $HST$/STIS imaging spectroscopy
suggest that the circular ring in the WFC3 images is
actually a ring plus caps or a 3-dimensional spherical shell expanding
with a velocity of 
approximately 
1000~km~s$^{-1}$ \cite{sokoloski2017a}.  Thus, the HST observations show 
that the equatorial torus that presumably shaped
the faster, bipolar flow is part of a more
complete, spherical core.  Interestingly, by three and a half years
into the eruption, H$\alpha$ (and [OIII]) emission from the outer,
fast flow had almost completely faded, leaving a remnant that was much
more spherically symmetric than the original, bipolar morphology.  We
identify this spherical core --- with its equatorial density
enhancement --- as the slow component of the ejecta.

Compared to the H$\alpha$ emission, [OIII] emission from V959~Mon
traces more diffuse gas.  Detectable levels of [OIII] emission emanated
from the same features as H$\alpha$ --- except for the dense, edge-on
equatorial torus (see Fig.~\ref{fig:v959monhst}; right side).
The extent of the [OIII] remnant in 2014 November (day 875 after
$\gamma$-ray detection) along the major axis was 1.05$''$ (measured from
locations that were 10\% of the peak flux).
The spherical shell and two knots to the east and west of the
central point source that were quite faint in H$\alpha$ outshined most
other features in [OIII].  And some hints of tails 
extending outward from the two [OIII] knots are also present. In 2014
December, STIS also detected Ne[V] ($\lambda$3426\AA) emission from the
regions of the [OIII] knots, whose emission was Doppler shifted in the
same sense as the larger lobes.  
As the [OIII] remnant expanded between 2014 and 2015, the inner
spherical shell took on the appearance of two arcs.  Even more so than
in the H$\alpha$ images, the outer lobes almost completely faded in
[OIII] by the 2015 November.  We describe the $HST$ WFC3 and STIS
observations of V959~Mon in detail in a forthcoming paper
\cite{sokoloski2017a}.

\begin{figure}
\begin{center}
\includegraphics[width=1.3\textwidth,angle=-90]{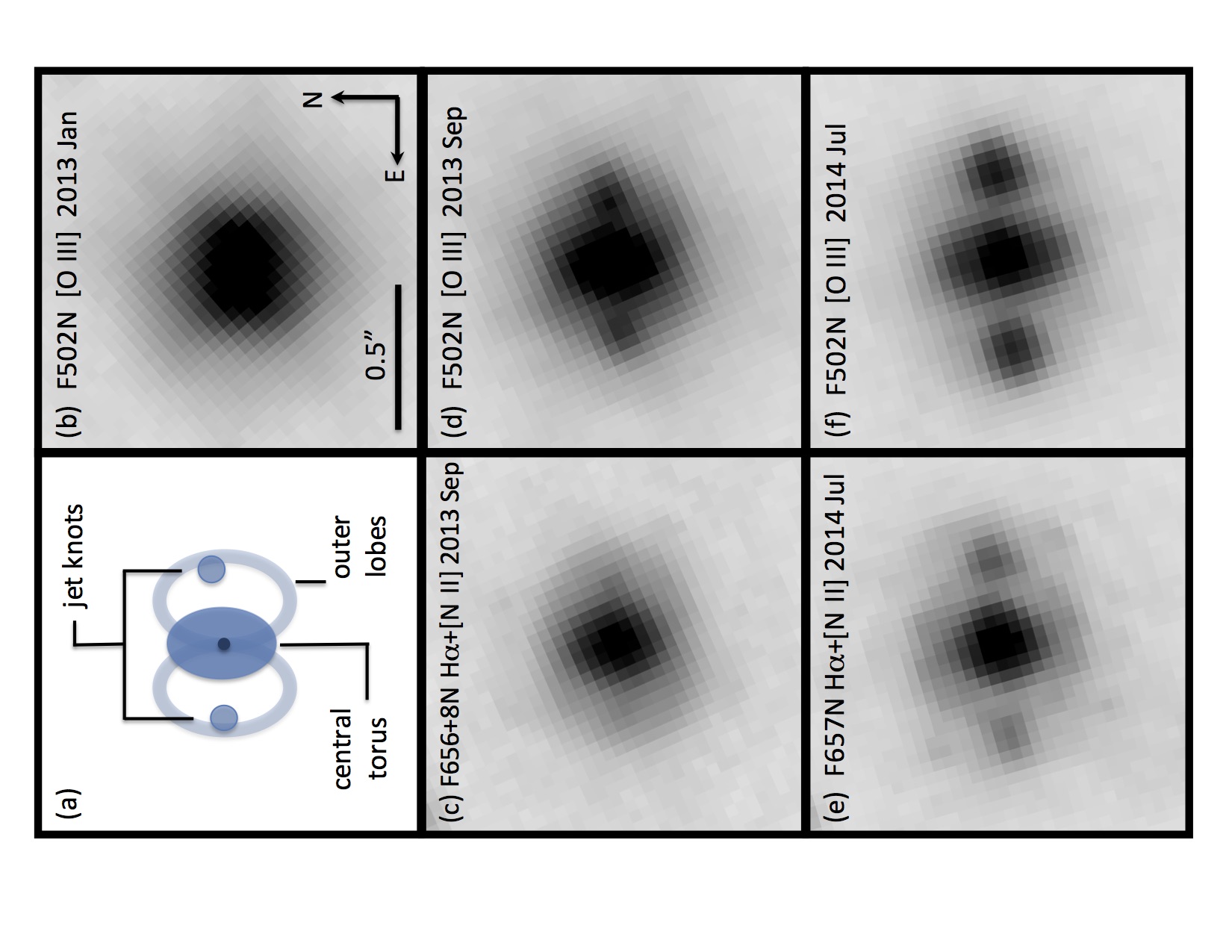}
\end{center}
\caption{$HST$/WFC3 imaging of T~Pyx.  All frames are 34 pixels = 1.3$''$
square, N up, with logarithmic intensity scaling and identical stretch
(pegged to brightest pixel in F502N image in each epoch/row). (a) A
to-scale schematic of the potential ejecta structures. (b) F502N 2013
January. (c) A sum of F656N+F658N 2013 September (d) F502N 2013
September.  This epoch best shows the outer lobes and first reveals
the jet knots.  (e) F657N 2014 July, this filter provides complete
velocity sampling of all components in H$\alpha+$[N II].  (f) F502N 2014
July.  This latest epoch shows significant proper motion in the jet
knots.
}
\label{fig:tpyxhst}
\end{figure}

{\bf $HST$ observations of T~Pyx.} Moving on to another nova-producing
binary, $HST$ observations of the 
recurrent nova T~Pyx highlight the similarities among even quite
disparate novae.  We observed T~Pyx with $HST$/WFC3 on 2012 July 12;
2013 January 30, June 19, July 15, and September 17-18; and 2014 July
10.  We also obtained 9 observations with $HST$/STIS (gratings G430L
and G750L) spanning this same time period (through
programs 12448, 13400, and 13796 (PI: Crotts)).  
Although T~Pyx is
generally considered to be a quite unique nova-producing binary, with
its short orbital period (of 1.8 hr; e.g. \cite{patterson2017})
and high rate of mass transfer \cite{knigge2000,uthas2010},
its ejecta
contain the same basic components as that of V959~Mon.  Light-echoes
and flash ionization in the well-known old remnant after the recent
nova eruption in 2011 demonstrated that its 3-dimensional shape is
dominated 
by an inclined torus
\cite{sokoloski2013,shara2015}.  And as with V959~Mon, long-slit
  spectroscopy of the old shell by \cite{obrien1998} suggests that the
  inclined torus is part of a more complete, spherical shell.  Once the ejecta from the 2011
eruption had had a few years to expand, $HST$/WFC3 images showed that they
also contained a central core consistent with an inclined torus
(Figure~\ref{fig:tpyxhst}; the central torus is most clearly visible
in the latest images, at the bottom of the figure); kinematics of the
central structure from STIS spectroscopy indicate that the torus is
inclined in the same direction as the torus in the old remnant and
expanding with a similar velocity (\cite{sokoloski2017b}, based on a
comparison with the 
proper motion of knots in the old remnant measured by \cite{schaefer2010}).
Moreover, our $HST$/STIS images 
show that, as in V959~Mon, the equatorial
torus appears to have mechanically shaped a faster flow into bipolar lobes
(although they are much fainter in T~Pyx than in V959~Mon).  The
lobes are not easily recognizable in $HST$ images of the old
remnant \cite{shara1997,schaefer2010}.  However, \cite{shara1989}
reported an ``extended envelope'' component (in H$\alpha\, +$~[NII])
in the remnant that could perhaps have been from faster, bipolar
lobes. 

Fig.~\ref{fig:tpyxhst} also shows that the ejecta from T~Pyx comprised
additional features besides the equatorial ring and bipolar lobes.
$HST$ resolved two bright knots that we will refer to as {\em jet
  knots} from 2013 June onward (in [OIII], although bipolar
structure was already evident in the spatially resolved STIS spectra
from 2012 July), within the eastern and western lobes.  Radial
velocities of approximately $\pm 800$~km~s$^{-1}$ from STIS
spectroscopy combined with proper motions (and taking d=4.8~kpc; \cite{sokoloski2013})
indicates that they moved away from the central
binary with a velocity of approximately 2000 km~s$^{-1}$.
They thus moved with a speed of more than
twice the average expansion speed of material in the
equatorial torus.  Furthermore, if the jet knots moved perpendicular
to the orbital plane, then the inclination of the binary must be on
the high side of reported range of values.
Moreover,
  the resemblance between the bright jet knots in T~Pyx and the
  relatively fainter knots within the lobes of V959~Mon supports
  conclusions from previous studies that nova remnants have other
  pervasive features in addition to equatorial rings and bipolar
  lobes.
We will describe the
  $HST$ WFC3 and STIS observations of T~Pyx in detail in a forthcoming
  paper \cite{sokoloski2017b}.

{\bf Secondary features.} Various imaging studies of nova remnants
have led to the determination 
that many remnants contain not only bipolar 
lobes and equatorial rings, but also `polar blobs', `polar
caps', `polar rings', and knots with tails.
Here we speculate
that these secondary features might all be related to the
main features in our $HST$ images of the young
(few-year-old) remnants of V959~Mon and T~Pyx --- if we take into
account the possibility that most of the ejecta mass
could be expelled somewhat later than
the initial 'puff' of material at $t_0$ (the time of the TNR).
Strong evidence for a delay of on the order of a month between the TNR
and the expulsion of most of the mass in both
V959~Mon and T~Pyx comes from the evolution of the radio and
X-ray emission from shock-heated plasma
\cite{nelson2014,chomiuk2014,linford2015}.  In T~Pyx, the delayed
ejecta clearly overtook and collided with slightly slower-moving
material (\cite{shore2011} detected P~Cyg absorption features during the first
with radial velocities of $-1000$ to
$-1500$~km~s$^{-1}$) about two months after the start of the eruption
\cite{nelson2014, chomiuk2014}.  As discussed by \cite{chita2008}, the
collision between an initial spherical shell and a later, faster
bipolar flow naturally leads to polar caps, blobs, or rings.  And some novae
experience even more than two episodes of mass ejection; 
V1369~Cen experienced a series of ejections, each expelling material
with velocities than the previous one (F.~Walter, private
communication).  In any 
  case, the features we call jet knots in T~Pyx, and the analogous
features in V959~Mon, could easily be described as polar blobs
(especially once the bipolar lobes faded).  Finally, with faster
flows overtaking slower flows at multiple locations within a given
remnant, there are plenty of opportunities for instabilities to
produce knots.

{\bf Implications of flow structure on $\gamma$-rays.} Observations
suggesting that the 
ejecta from novae fundamentally 
consist of two colliding flows --- a slow flow with an equatorial
density enhancement and a faster flow that is shaped by the slow flow
--- have bearing on $\gamma$-ray production and mass ejection in these
events.  For starters, if the V959~Mon scenario of
$\gamma$-ray production is widespread, then $\gamma$-rays are
expected to be strongest from systems with slow (less than
approximately 1000~km~s$^{-1}$), radiative shocks.  Such shocks in
novae have high enough densities and low enough expansion speeds to
efficiently trap relativistic protons, which then collide with
non-relativistic protons to produce pions that decay and emit
$\gamma$-rays \cite{metzger2014,metzger2015}.  Work by
\cite{metzger2015} showed this so-called {\em hadronic} scenario for
the production of $\gamma$-rays in novae to be favored over the {\em
  leptonic} scenario of inverse Compton scattering by relativistic
electrons.  They argued that because the neutral material ahead of the
forward shocks can efficiently reprocess X-rays from shock-heated
plasma into the optical (at the early times when $\gamma$-rays are
being produced), the ratio of $\gamma$-ray to optical fluxes places
constraints on the efficiency of particle acceleration. Such
constraints for both V1324~Sco and V339~Del were quite high and
therefore more consistent with hadronic than leptonic scenarios.
Other surprising results from the models explored by
\cite{martin2013,metzger2014,metzger2015,vlasov2016,vurm2016} are that the
observed $\gamma$-ray luminosities require shocks in novae to be very
powerful (with power comparable to the Eddington luminosity for a
white dwarf), and that shocks probably 
convert the kinetic energy of the fast flow into
optical emission for
at least a few weeks around maximum optical light.

{\bf Common-envelope interaction.} In addition, by highlighting the
importance of material concentrated in the orbital plane, recent
investigations of $\gamma$-rays from novae have resurrected past
questions about the role of the donor star in unbinding the WD
envelope after the TNR (see, e.g., \cite{chomiuk2014}).  In other
words, is common-envelope evolution important in shaping and/or
ejecting the remnants of classical novae (e.g., \cite{livio1990})?
Whether or not common-envelope effects are at play in classical novae,
this mechanism is unlikely to be important for wide, symbiotic
binaries because their donor stars are far from their eruptive WDs.
So, does the {\em lack} of interaction between the expanded WD
photosphere and its binary companion lead to lower ejection efficiency
and therefore lower ejecta masses in
some wide, symbiotic binaries?  If so, could this difference help
explain how quasi-steady shell burning can persist for up to centuries
after novae in some symbiotic stars (such as AG Peg
\cite{ramsay2016,tomov2016}).
Regardless of
the answers to these questions, the flow structure, $\gamma$-ray
production, and ejection of the WD envelope seem to be strongly
linked.

{\bf Pre-maximum halts, THEAs, and dust.} Finally, new
understanding of the flow structure and shocks in novae may also have
implications for the shape of optical light curves (including
pre-maximum halts), the transient heavy element absorbing gas (THEA
features in optical spectra; \cite{williams2008}), and the formation of dust.
Although \cite{hillman2014} propose
the pre-maximum halts are due to changes in
convective energy transfer, and this effect might well
have an impact on optical light curves, the multi-component flow structure
also
naturally leads to a `pre-maximum halt' (as is sometimes observed
in optical light curves; \cite{hounsell2010}).
If the initial optical rise is due to the expansion of an 
optically thick photosphere near the outermost edge of the ejecta, then the
optical light curve 
must turn over or halt as this material becomes optically thin.  During the
halt, the optical (pseudo-)photosphere recedes with respect to the
outer edge of the expanding ejecta.  When it reaches the slow, dense
core, with its greater optical depth, the optical brightness would be
  expected to rise a second time along with the increase in
size of the expanding core (especially if the optical brightness of the
core is high due to shock heating).  The optical brightness peaks
when the core reaches its maximum size before beginning to
become optically thin.  Near optical maximum, one might expect to see
blueshifted absorption by the material in the (slow) core along with
emission from the optically thin fast flow.   Such absorption
features, with blueshifted velocities of several hundred km~s$^{-1}$,
have been observed and described as THEA lines (by
\cite{williams2008}).
With
regard to the long-standing question of how dust can form in the harsh
environment of a nova, \cite{derdzinski2016} have recently shown that
the dense, cool gas behind the type of radiative shock needed for
$\gamma$-ray production also provides an ideal environment for the
creation of dust grains.  Thus, several long-standing
problems could be resolved once we take into account the structure and
microphysics of colliding
flows within the ejecta from novae.

\section{Conclusions} 

\subsection{Symbiotic stars}

\begin{itemize}

\item{We define a WD symbiotic star as a binary in which a red giant
  transfers enough material to a WD for the interaction to produce an
  observable signal at some waveband.}

\item{The presence or absence of quasi-steady shell burning sets the
  luminosity of the accreting WD in a symbiotic and affects its
  appearance across the electromagnetic spectrum.  We refer to
  symbiotic stars with and without shell burning on the surface of
  their WDs as {\em burning} and {\em non-burning} symbiotic stars,
  respectively.}

\item{Non-burning symbiotic stars are difficult to find in optical
  spectroscopic surveys.  Because their WD masses, mass transfer
  rates, space densities, and other properties are likely to differ
  from those of burning symbiotics, alternative types of searches are
  needed to draw accurate conclusions about the nature and evolution
  of this class of wide binaries.}

\item{Non-burning symbiotics can be identified in the X-rays, by
  their UV excess, and by their UV variability.}

\item{Non-burning symbiotics offer a clearer view of the accretion
  flow from the red giant to the WD than burning symbiotics.}

\end{itemize}

\subsection{Novae}

\begin{itemize}

\item{The pervasive production of $\gamma$-ray emission shows that
  shocks are common and important during nova eruptions whether they
  are embedded or not. The emission of $\gamma$-rays by non-embedded
  (classical) novae requires complex outflows that
  result in internal shocks.}

\item{The basic structure of ejecta consists of a core plus halo.
  Early images of V959 Mon and T Pyx share common characteristics of
  equatorial rings and bipolar lobes that extend to larger angular
  sizes, even though these novae are dissimilar in many ways. It is
  likely that the core has an equatorial enhancement, which interacts
  with the faster flow (halo) and shapes it bipolar lobes, not only in
  these novae but most others.}

\item{By three and a half years after the start of the eruption, the
  fast component of the V959~Mon had faded almost beyond
  detectability. Images of older nova shells might show primarily the
  slow component, which is more massive and denser than the faster
  component.}

\item{Collisions between the slow-moving material that comprises the
  core and the faster flow produce strong shocks that are crucial for
  accelerating particles and generating gamma-rays. The same shocks
  produce non-thermal radio emission, thermal X-rays, and perhaps a
  large fraction of optical light, at different phases of the
  eruption.}

\item{Now that they have been recognized as $\gamma$-ray sources, novae
  serve as unique laboratories for the study of particle
  acceleration. Multiwavelength observations contemporaneous with
  $gamma$-ray detection have the potential to allow us to distinguish
  between leptonic and hadronic processes, and to constrain the shock
  properties, including the efficiency of particle acceleration.}

\end{itemize}

\acknowledgments

We are grateful for conversations with J.~Miko{\l}ajewska,
T.~Nelson, G.~J.~Luna, A.~Lucy, M.~Rupen, L.~Chomiuk, and the other
members of the ENova collaboration.  G.~J.~Luna and A.~Lucy, and gave
helpful comments on this manuscript.  The authors acknowledge support
from HST grants GO-13400, GO-13796, and GO-13715, as well as NNX15AF19G
and AST-1616646.

\end{document}